# A coupled FE-BE approach for vibro-acoustic response prediction of laminated composite panels due to turbulent boundary layer excitation involving Cholesky decomposition

**Biplab Ranjan Adhikary[1]\*, Atanu Sahu[2], Partha Bhattacharya[3]**

[1,3]Jadavpur University, (Department of Civil Engineering), Kolkata, (West Bengal), India 700 032

[2]National Institute of Technology Silchar, (Department of Civil Engineering), Silchar, (Assam), India 788 010

\*corresponding author, e-mail: biplab.iitkgp@yahoo.com

*An original numerical framework is developed in the present research work in order to estimate the free field sound radiation from baffled structural panels subjected to turbulent boundary layer (TBL) flow-induced excitation. A semi-analytical method is used to estimate the TBL wall pressure spectrum which is decomposed using Cholesky's technique to obtain random wall pressure in the frequency domain. Structural panels are modeled using the finite element technique and a coupled finite element-boundary element modeling technique is developed to estimate the sound power level radiating into the free field. Results are obtained for laminated composite structural panels with various fiber orientations and significant findings are discussed. The developed technique has the potential to be further extended for complex structures in terms of geometry, material properties, and boundary conditions. The complete numerical toolbox, developed in an in-house MATLAB environment, enables the prediction of turbulent-structure acoustic coupled behavior at an early design stage.*

## 1  Introduction

Turbulent boundary layer (TBL) induced structural vibration and resulting sound radiation has been a major area of engineering research over years. The two important areas where the TBL-induced acoustic radiation is a major cause of concern are aircraft and surface transport vehicles. Furthermore, in recent years the use of laminated composites in vehicular structures which are inherently lightweight with low damping has necessitated the importance to understand the sound radiation mechanism in the vehicular cabin for passenger comfort. The primary source of sound generation within an aircraft interior is skin panel vibration induced due to unsteady boundary layer flow, jet propulsion noise, propeller noise, and noise generated due to control surface operation. There has been a significant improvement in engine technology to make power plants 'quieter'. Similarly, in the automobile industry, efforts are being made to reduce sound transmission from the engine compartment and the sound generated from tire-road interaction. As a result, the major cause of concern that still remains in terms of noise transmission into the passenger compartment is the TBL-induced noise.

Since the early 1960's with the rapid progress in jet propulsion technology and with the introduction of long-haul flight, passenger comfort in terms of noise level within the passenger cabin has been a major focus area in aeronautical research. One of the earliest demonstrations of the effect of jet noise in the aft of a passenger cabin during climb condition was presented by Bishop [1]. In an experimental study on a real aircraft (Convair 880) in cruise condition, he demonstrated the significant contribution of boundary layer excitation in cabin noise over a wide frequency range. Experimental flight test studies by various researchers [2, 3, 4] have shown that there is a correlation between the interior cabin acoustic pressure with the flow speed due to the coupling between the fuselage structure and the boundary layer flow. An experimental study on real-life flight vehicles is always an expensive idea in terms of time, manpower, and financial resources. Hence, attempts are made by several researchers to develop a theoretical framework for the study of sound generation from flexible panels subjected to excitation due to turbulent boundary layer excitation.

The theoretical background needed to fully appreciate the sound generation mechanism from vibrating flexible panels due to TBL excitation can be classified into two distinct mathematical processes – (1) Turbulent boundary layer modeling and TBL-Structure interaction and (2) Structural – Acoustic modeling for estimating sound radiation from vibrating panels. Although research articles dealing with each of the above areas are in abundance there is only a handful of them available wherein a coupled model is presented.

One of the earliest attempts to estimate turbulence-induced wall-pressure fluctuation can be attributed to Kraichnan [5]. Subsequently, Lowson [6] derived the governing equations for the pressure fluctuations in TBL from known empirical formulations of RMS pressure fluctuations, obtained from experimental works and referring to incompressible Navier-Stokes equations. Thereafter estimations of pressure fluctuations were made through empirical and semi-empirical formulations by many researchers. Mean-square pressure fluctuations which essentially is a measure of total energy due to TBL pressure fluctuation were well estimated by Lilley et al. [7], Corcos [8, 9], Bull [10], Blake [11], Schewe [12], Farabee and Casarella [13] and others. Some of the famous single-point wall-pressure spectrum (normalized wavenumber-frequency spectrum) models were proposed by several researchers or research groups. Efimtsov [14] proposed a model considering the dependency of pressure-spectrum on Mach number (M), Reynolds number (Re), and Strouhal number (Sh). The model was based on a series of flight test data in a wide range of Mach Numbers and Reynolds numbers. Goody [15] proposed an empirical model for single-point wall-pressure spectra beneath a two-dimensional, zero pressure gradient boundary layer which was based on the experimental surface pressure spectra measured by seven research groups. This model received wide acceptance among researchers because of its applicability in quite a large range of frequencies and Reynolds number. Smol'yakov and Tkachenko [16] identified different scaling of wall-pressure spectrum for different frequency regimes and formulated accordingly.

One of the earliest numerical models available in the open literature to estimate TBL flow-induced structural response can be attributed to Hambric et al [18]. In their method, they used the Smol'yakov and Tkachenko model [16] to obtain wall-pressure spectrum at a single point on a plate placed on the infinite baffle. The spatial coherence function proposed by Corcos [8] was used to obtain cross power spectrum of wall-pressure fluctuations at different points. This cross-power spectrum when coupled with the frequency response function (or transfer function matrix) of the plate structure resulted in structural response in terms of displacement and velocity.

Literature on the treatment of sound generation processes due to TBL flow-induced excitation within a single mathematical framework is very limited. A series of articles by Rocha and her co-workers [19, 20] can be regarded as one of the earliest research contributions wherein the sound generation mechanism due to TBL flow-induced excitation is presented within a single analytical framework. However, the limitation of the framework presented therein is that it is limited to structures with simple geometry and boundary conditions. Recently, Caiazzo et al [21] presented a fully coupled structural-acoustic model in an analytical framework to estimate and control sound radiation into a double-walled cavity subjected to a TBL flow-induced excitation. A semi-empirical model defined by Goody [15] is used to obtain the single point wall pressure spectrum and the Corcos model is used to calculate the cross-spectral density. Modal expansion functions for in-vacuo structural modes and rigid wall acoustic cavity modes are used to develop the coupled structural-acoustic behavior. Due to its theoretical treatment, the method is limited to simply supported boundary conditions for the structure and for simple geometrical configuration.

The present research work is therefore an attempt to overcome the modeling limitation as presented by the earlier researchers. A fully coupled numerical model in a finite element (FE) – boundary element (BE) framework is presented in this article to address the problem of the coupled structural-acoustic behavior with the structure subjected to a TBL flow. A semi-empirical single point wall pressure spectrum model as defined by Smolyakov and Tkachenko [16] is used and a cross-spectral density matrix is developed using the spatial coherence function proposed by Corcos [8]. The cross-power spectral density is decomposed using Cholesky's technique to estimate random pressure signal in the frequency domain at all the grid points of the baffled orthotropic panel. Subsequently, the fully coupled FE-BE model is used to calculate the free field radiated sound power from the other side of the plate. A schematic of the physical domain conceived is shown in Figure 1 and the detailed mathematical derivation is presented in the next section.

## 2 Mathematical Model

The mathematical description of the physical problem consists of three (3) subsections – (1) TBL model (2) FE model description of the orthotropic panel to estimate the panel vibration and (3) BE model to obtain the radiated sound power in the free field. The detailed formulation of each component is presented below -

### 2.1 Turbulent Boundary Layer Pressure Spectrum Model

In the present work, a low Mach number flow is considered and the frequency regime considered is between 0 – 500 Hz.

An empirical formula to calculate turbulent boundary layer displacement thickness is reported in the doctoral thesis of Mahmoudnejad [17] as

$$\delta^* = \frac{0.0174x}{\text{Re}_x^{0.139}} \qquad (1)$$

As the present study assumes homogeneous and fully developed turbulence over the plate, the above-mentioned formula is used to calculate the boundary layer displacement thickness which is subsequently used in the single-point wall-pressure spectrum calculation, using Smol'yakov and Tkachenko [16] model as given in Eq. (2)

$$\Phi_p(\omega) \approx \left(\frac{\tau_w \delta^*}{U_0}\right)\left(\frac{5.1}{1 + 0.44\left(\frac{\omega \delta^*}{U_0}\right)^{\frac{7}{3}}}\right) \quad (2)$$

where $U_0$ is the free-stream flow velocity and $\delta^*$ is the boundary layer displacement thickness. In the case of TBL flow with zero pressure gradient wall shear stress $\tau_w$ can be estimated as $\tau_w \approx 0.0225\rho U_0^2/Re_\delta^{0.25}$, where $Re_\delta$ is the boundary layer thickness Reynolds number given by $Re_\delta \approx 8U_0\delta^*/\nu$, with $\nu$ being the kinematic viscosity of the fluid, and $\rho$ the density of the fluid. Once the single point power spectrum is obtained, cross-spectra of the pressure fluctuations are obtained in similar in lines to that presented by Hambric *et al.* [18]. A modified Corcos model [8] is adopted to obtain the correlation function $\Gamma(\xi_1, \xi_3, \omega)$ which is multiplied with the single point power spectrum of pressure fluctuations to obtain cross power spectrum of different points, as a function of frequency and separation vectors along both the streamwise and cross-flow direction. Cross-spectrum is calculated as,

$$\Phi_{pp}(x_\mu, x_\nu, \omega) = \sqrt{\Phi_p(x_\mu, \omega)\Phi_p(x_\nu, \omega)}\,\Gamma(\xi_1, \xi_3, \omega) \quad (3)$$

The spatial correlation function originally proposed by Corcos [8] is

$$\Gamma(\xi_1, \xi_3, \omega) = A_1\left(\frac{\omega\xi_1}{U_c}\right) B_1\left(\frac{\omega\xi_3}{U_c}\right) \quad (4)$$

with $\xi_1$ and $\xi_3$ are separation vectors between two points, along x (stream-wise) and z (cross-flow) direction. $U_c$ is the average convective velocity approximated by Bull [10] as

$$U_c \cong U_0\left(0.59 + 0.30 e^{-\frac{0.89\omega\delta^*}{U_0}}\right) \quad (5)$$

$$A_1\left(\frac{\omega\xi_1}{U_c}\right) = \left(1 + \alpha_1\left|\frac{\omega\xi_1}{U_c}\right|\right) e^{-\alpha_1\left|\frac{\omega\xi_1}{U_c}\right|} e^{i\frac{\omega\xi_1}{U_c}} \quad (6)$$

$$B_1\left(\frac{\omega\xi_1}{U_c}\right) = e^{-\alpha_3\left|\frac{\omega\xi_3}{U_c}\right|} \quad (7)$$

where $\alpha_1$ and $\alpha_3$ are decay constants. For spatially homogenous pressure field $\alpha_1$ and $\alpha_3$ are considered as 0.11 and 0.70 (Hambric *et al.* [18]).

## 2.2 FE modeling of Orthotropic Laminate

A 4-node iso-parametric quadrilateral element with five degrees of freedom per node is used to discretize the structural panel. As per the standard procedure, the element stiffness matrix **[K]** and the element mass matrix **[M]** for a 2-D element can be written as,

$$[K] = \int_0^a \int_0^b [B]^T [D][B] dx dy \qquad (8)$$

and,

$$[M] = \int_0^a \int_0^b [N]^T [\rho][N] dx dy \qquad (9)$$

The layered fiber reinforced plastic lamina introduces coupling between in-plane, out of plane and shear behavior in the laminated panel which is modeled through the **[D]** matrix. The **[D]** matrix is given by,

$$[D] = \begin{bmatrix} A_{11} & A_{12} & A_{16} & B_{11} & B_{12} & B_{16} & 0 & 0 \\ A_{12} & A_{22} & A_{26} & B_{12} & B_{22} & B_{26} & 0 & 0 \\ A_{16} & A_{26} & A_{66} & B_{16} & B_{26} & B_{66} & 0 & 0 \\ B_{11} & B_{12} & B_{16} & D_{11} & D_{12} & D_{16} & 0 & 0 \\ B_{12} & B_{22} & B_{26} & D_{12} & D_{22} & D_{26} & 0 & 0 \\ B_{16} & B_{26} & B_{66} & D_{16} & D_{26} & D_{66} & 0 & 0 \\ 0 & 0 & 0 & 0 & 0 & 0 & A_{44} & A_{45} \\ 0 & 0 & 0 & 0 & 0 & 0 & A_{45} & A_{55} \end{bmatrix} \qquad (10)$$

The coefficients are given by

$$A_{ij} = \sum_{k=1}^{n} (Q'_{ij})_k (z_k - z_{k-1})$$

$$B_{ij} = \frac{1}{2} \sum_{k=1}^{n} (Q'_{ij})_k (z_k^2 - z_{k-1}^2)$$

$$D_{ij} = \frac{1}{3} \sum_{k=1}^{n} (Q'_{ij})_k (z_k^3 - z_{k-1}^3); \quad i,j = 1, 2, 6$$

Following the first order shear deformation theory as proposed by Reissner [22] and Mindlin [23], a shear correction factor, $k_s$ is introduced in the constitutive relation for transverse shear. Therefore, the coefficients are given by,

$$A_{ij} = k_s \sum_{k=1}^{n} (Q'_{ij})_k (z_k - z_{k-1}); \quad i,j = 4, 5 \text{ and} k_s = \frac{5}{6}$$

The on-axis constitutive matrix coefficients are given as follows –

$$Q_{11} = \frac{E_1}{1 - \nu_{12}\nu_{21}}; Q_{12} = \frac{\nu_{12}E_2}{1 - \nu_{12}\nu_{21}}; Q_{22} = \frac{E_2}{1 - \nu_{12}\nu_{21}}; Q_{66} = G_{12}; Q_{44} = G_{23} \text{ and } Q_{55} = G_{13}$$

The inertia matrix $[\rho]$ is expressed as,

$$[\rho] = \begin{bmatrix} I_1 & 0 & 0 & I_3 & 0 \\ 0 & I_1 & 0 & 0 & I_3 \\ 0 & 0 & I_1 & 0 & 0 \\ I_3 & 0 & 0 & I_2 & 0 \\ 0 & I_3 & 0 & 0 & I_2 \end{bmatrix} \quad (11)$$

with, $\{I_1, I_2, I_3\} = \int_{-h/2}^{h/2} \rho(z)\{1, z, z^2\}dz$

Once the elemental matrices are obtained, an assembling procedure is carried out to obtain the global stiffness and the global mass matrix and thereafter they are suitably used to obtain the FE governing equilibrium equation for the flexible structural panel subjected to the external disturbances, $F_{tbl}$, in the form of TBL flow induced pressure fluctuation. The governing equation can thus be given as

$$[M_s]\{\ddot{d}_s\} + [C_s]\{\dot{d}_s\} + [K_s]\{d_s\} = \{F_{tbl}\} \quad (12)$$

## 2.3 Boundary element formulation and structural-acoustic coupling

As the panel vibrates due to the TBL-induced excitation, the structure perturbs the acoustic medium adjacent to it and as a result, the energy is dissipated in the form of sound into the other half-space. The governing equation that mathematically describes the acoustic domain is the three-dimensional Helmholtz equation as shown below,

$$(\nabla^2 + k^2)p = 0 \quad (13)$$

Using the advantage of the boundary element formulation, the three-dimensional acoustic domain is reduced to the surface problem over the panel only by applying Gauss' second identity,

$$\int_v (p\nabla^2 g - g\nabla^2 p)dV = \int_S \left(p\frac{\partial g}{\partial n} - g\frac{\partial p}{\partial n}\right)dS \quad (14)$$

Since, the panel radiates the sound into an infinite half-space, there is no reflection of the acoustic energy from the infinite boundary. This is ensured by applying the Sommerfeld radiation conditions for the external acoustic problem [24]. Considering the momentum balance between the structure and the adjacent fluid particles, the final boundary integral equation is written as follows;

$$C(X)p(X) - \int_S p(Y)\frac{\partial g(X,Y)}{\partial n}dS(Y) = \int_S j\rho\omega \dot{d}_s(Y)g(X,Y)dS(Y) \quad (15)$$

where, $p(X)$ is the sound pressure radiated by the panel at a point due to the structural panel vibrating with a velocity $\dot{d}_s(Y)$ and $g(X,Y)$ is the free-space Green's function. $C(X)$ is a geometrical constant, which makes it possible to apply the equation for any structure having complex geometry and/or boundary conditions.

## 3 Solution Procedure

The solution process to the entire mechanism of sound generation from infinitely baffled panels subjected to TBL flow induced excitation is based on the assumption that statistically the wall pressure fluctuation beneath the TBL flow is homogeneous, stationary and Gaussian. In order to obtain vibration response from the panels subjected to stochastic excitation like TBL pressure fluctuations, several researchers, e.g., Hambric *et al*. [18], Rocha [19], and others have followed the convention of Bendat and Piersol [25] and Lin [26]. In this technique, the output power spectral density is obtained in terms of a cross power spectral density matrix as a function of frequency. Rocha [20] in their article extended this idea to obtain the radiated sound power by taking the product of the output velocity spectra with the radiation resistance following the work by Wallace [27].

In the present article, a new technique to obtain vibration response is proposed based on the work by Wittig and Sinha [28]. In a seminal work, Wittig and Sinha [28] showed that for any Gaussian random process for which Cross Spectral Density function $G_{\nu\mu}(\omega)$ between two random time processes $x_\nu$ and $x_\mu$ is available it can be factored into a lower triangular matrix $[L_{\nu\mu}(\omega)]$ and its complex transpose such that

$$[G_{\nu\mu}(\omega)] = [L_{\nu\mu}(\omega)][L^*_{\nu\mu}(\omega)]^T \tag{16}$$

where * denotes the complex conjugate and $\nu, \mu = 1, 2, \ldots, M$ are discrete points.

If one considers a random time series with time interval $h$ having a total number of $N$ points the Fourier transform pair will have a frequency interval of $1/Nh$. Now if a Fourier transform pair exist such that $x_\nu$ and $X_\nu$ are Fourier transform pair, they can be related by

$$x_p(nh) = \frac{1}{N}\sum_{k=0}^{N-1} X_p(k/Nh) exp\left(j\frac{2\pi kn}{N}\right) \tag{17}$$

and

$$X_p\left(k/Nh\right) = \sum_{n=0}^{N-1} x_p(nh) exp\left(-j\frac{2\pi kn}{N}\right) \tag{18}$$

The term $X_p$ can then be expressed in the matrix form

$$\begin{Bmatrix} X_1 \\ X_2 \\ \vdots \\ \vdots \\ X_M \end{Bmatrix} = \left(\frac{N}{2h}\right)^{1/2} \begin{pmatrix} L_{11} & 0 & \cdots & 0 \\ L_{21} & L_{22} & \cdots & 0 \\ \vdots & \vdots & \vdots\vdots\vdots & \vdots \\ L_{M1} & L_{M2} & \cdots & L_{MM} \end{pmatrix} \begin{Bmatrix} \zeta_{1k} \\ \zeta_{2k} \\ \vdots \\ \vdots \\ \zeta_{Mk} \end{Bmatrix} \tag{19}$$

In the present formulation the lower triangular matrix $[L_{ij}]$ is obtained through Cholesky decomposition of the wall-pressure cross-spectrum $\Phi_{pp}$, with $\zeta_{ik}$ as independent Gaussian random number set having mean = 0 and variance = 0.5. The vector $\{X_\nu\}$ can thus be regarded as the random TBL induced pressure fluctuation which can be related to forcing function in the frequency domain $\{F(\omega)\}$ by a suitable mapping matrix [R]

$$\{F(\boldsymbol{\omega})\} = [R]\{X_v\} \tag{20}$$

Using suitable modal transformation and converting the same into frequency domain, the structural response equation given in Eq. (12) in the modal domain can be written as

$$\{q(\omega)\} = [H(\omega)]\{F(\omega)\} \tag{21}$$

where, $[H(\omega)]$ is the frequency response function defined by

$$H(\omega) = \frac{\varphi \, \varphi^T}{\bar{m}(-\omega^2 + 2\xi\omega_n\omega + \omega_n^2)} \tag{22}$$

with $\varphi$ being the eigen vector (mode shape) and $\bar{m}$ the modal mass.

Subsequently, the modal displacement $q(\omega)$ is transformed into the nodal domain using mode summation procedure and expressed as $d_s(\omega)$.

Once the vibration responses are obtained for all the points on the panel and the velocities in the frequency domain are calculated, one can invoke Eq. (15) to obtain the radiated sound pressure.

The estimation of the energy transmission behavior of the vibrating plates due to TBL induced excitation is expressed in terms of two parameters. They are

a) Average Quadratic Velocity $\langle \mathbf{V}^2 \rangle$

$$\langle \mathbf{V}^2 \rangle = \frac{\omega^2}{2A} \int_A \dot{d}_s(\omega)\dot{d}_s^*(\omega)dA \tag{23}$$

The average quadratic velocity thus obtained is finally expressed in dB scale referenced to 2.5 x $10^{-15}$ (m/s)$^2$

b) Average Radiated Sound Power Level

$$L_{p,rad} = 10\log(\langle P_{rad} \rangle / P_{ref}) \tag{24}$$

where, $\langle P_{rad} \rangle$ is the sound power averaged over the radiating plate surface given by,

$$\langle P_{rad} \rangle = \frac{1}{2} Re \int_A \dot{d}_s^*(\omega)p(\omega)dA \tag{25}$$

and the reference sound power level P$_{ref}$ is taken as $10^{-12}$ Watt.

In the present work the finite element modeling of the structural panel is carried out in Ansys (ver. 14.5) simulation package to obtain the structural modal parameters. The boundary element code used to model Eq. (15) is developed in-house using MATLAB (ver. R2013b) platform. The other necessary mathematical processes described are simulated using MATLAB. In the next section results are presented first to validate the developed technique. Subsequently, the vibration and acoustic responses for isotropic and orthotropic plates are obtained and are presented. The entire work flow is presented in Figure 2.

## 4 Results and Discussion

This section contains results for validating the present formulation with those reported in the open literature. Subsequently, case studies are simulated to estimate sound transmission behavior of orthotropic laminated plates subjected to TBL excitation.

Finite element analysis results as reported in Hambric *et al.* [18] is used to compare the velocity power spectral density of a vibrating steel plate with clamped condition on all the four sides. A schematic of which is shown in Figure 1. The modulus of elasticity of the plate is taken as 210 GPa, density is taken as 7800 kg/m$^3$, and a damping ratio of 0.0025 is considered. The length, $L_x$ and the width, $L_y$ of the plate is taken as 0.47 m and 0.37 m, respectively with the thickness taken as 1.59 mm. The wind flow speed is taken as 44.7 m/s. The plate is discretized using a 60 x 40 mesh. The velocity PSD of the plate at point 'A' (0.15m, 0.25m; Refer Figure 1) due to the TBL flow is obtained and shown in Figure 3. The spectral density plot compares well with those reported by Hambric *et al.* [18]. At higher frequencies (> 300 Hz), the power spectral density values from the present studies are little higher than those reported by Hambric *et al.,* and also there is a slight shift in the resonant frequencies. In the present analysis, although results are obtained up to 500 Hz, the structural transfer function is developed using the first 40 structural modes up to 2500 Hz. This might have caused a higher value in the spectral density especially at higher frequencies due to the modal contribution of the higher modes. It must be noted that in the present formulation the boundary layer displacement thickness, $\delta^*$ is calculated using the formula reported in the doctoral thesis of Mahmoudnejad [17], whereas, in the work reported by Hambric *et al.* [18] the displacement thickness taken was that from experimental studies. This might also have contributed to the slight discrepancies in the PSD values.

In the next phase of the work, the developed model is used to obtain the average quadratic velocity and the radiated sound power for isotropic and antisymmetric orthotropic panels subjected to TBL excitation due to a flow velocity of 44.7 m/s. The panels considered for the analysis are having dimensions $L_x$ as 0.5 m, and $L_y$ as 0.35 m (refer Figure 1) with 2 mm thickness having simply supported boundary conditions on all along the four edges. The left edge with 0.35 m width is considered as the inlet for the flow (Figure 3). Seven (7) different cases are considered for the analysis and they are as follows – Case I – Aluminum; Case II – Carbon Fiber Reinforced (CFRP) Laminated composite with (0/90) lamination sequence; Case III – CFRP laminated composite with (0/90/0/90) lamination; Case IV - CFRP laminated composite with (30/-30) lamination; Case V - CFRP laminated composite with (30/-30/30/-30) lamination; Case VI - CFRP laminated composite with (45/-45) lamination; Case VII - CFRP laminated composite with (45/-45/45/-45) lamination.

Material properties used to model the plates are as follows –
Aluminium – Young's Modulus (E) = 70 GPa, Poisson's ratio = 0.3, Density (ρ) = 2700 kg/m$^3$
CFRP laminates - $E_{11}$ = 138 GPa, $E_{22}$ = $E_{33}$ = 6.9 GPa. $\nu_{12}$ = 0.31, $\nu_{23}$ = $\nu_{13}$ = 0.3. $G_{12}$ = $G_{13}$ = 4.5 GPa, $G_{23}$ = 4.05 GPa. Density (ρ) = 1570 kg/m$^3$
Primarily the structural frequencies and the mode shapes for the panels are obtained and they are listed in Table 1. The average quadratic velocity in dB and the average radiated sound power level (SPL) in dB are then obtained following Eq.22 and Eq. 23, respectively. The average quadratic velocity and the average radiated SPL for the Aluminum panel is shown in Figure 4(a) and Figure 5(a), respectively. The average quadratic velocities for the orthotropic panels are presented in Figures 4(b), 4(c) and 4(d) while those for the average quadratic SPL are shown in Figures 5(b) – 5(d). The SPL in dB for the first few modes obtained for different cases are also listed in Table 2.

**Table 1 Frequency (in Hz) and mode numbers for simply supported rectangular panels having dimension 0.5 m x 0.35 m x 0.002 m**

| Case I | Case II | Case III | Case IV | Case V | Case VI | Case VII |
|---|---|---|---|---|---|---|
| 58.43 (1,1) | 52.96 (1,1) | 58.55 (1,1) | 50.11 (1,1) | 60.78 (1,1) | 55.57 (1,1) | 69.75 (1,1) |
| 116.11 (2,1) | 87.08 (2,1) | 111.13 (2,1) | 106.63 (2,1) | 137.36 (1,2) | 101.46 (2,1) | 136.34 (2,1) |
| 176.45 (1,2) | 134.63 (1,2) | 191.72 (1,2) | 109.97 (1,2) | 141.10 (2,1) | 134.73 (1,2) | 181.90 (1,2) |
| 212.52 (3,1) | 164.67 (3,1) | 220.45 (3,1) | 178.80 (2,2) | 237.47 (2,2) | 167.14 (3,1) | 231.14 (3,1) |
| 234.00 (2,2) | 165.09 (2,2) | 222.08 (2,2) | 192.36 (3,1) | 256.11 (1,3) | 196.64 (2,2) | 271.98 (2,2) |
| 330.21 (3,2) | 230.01 (3,2) | 302.20 (3,2) | 207.93 (1,3) | 266.01 (3,1) | 257.12 (4,1) | 357.75 (4,1) |
| 348.00 (4,1) | 264.86 (4,1) | 376.70 (4,1) | 274.61 (3,2) | 371.40 (2,3) | 265.54 (1,3) | 362.99 (1,3) |
| 374.29 (1,3) | 307.19 (1,3) | 427.30 (1,3) | 280.50 (2,3) | 376.93 (3,2) | 277.02 (3,2) | 391.56 (3,2) |
| 431.68 (2,3) | 325.76 (4,2) | 439.38 (4,2) | 314.54 (4,1) | 421.43 (1,4) | 330.44 (2,3) | 464.97 (2,3) |
| 465.45 (4,2) | 329.82 (2,3) | 447.99 (2,3) | 342.81 (1,4) | 437.99 (4,1) | 372.03 (5,1) | --- |

**Table 2 Average radiated SPL (in dB) from rectangular simply supported 0.5 m x 0.35 m x 0.002 m panels placed in a baffle subjected to a flow velocity of 44.7 m/s**

| Case | Case I | Case II | Case III | Case IV | Case V | Case VI | Case VII |
|---|---|---|---|---|---|---|---|
| 1st mode | 31.96 | 48.48 | 38.58 | 42.84 | 37.51 | 40.62 | 42.35 |
| 2nd mode | 30.36 | 25.52 | 28.98 | 26.68 | 30.09 | 20.77 | 31.92 |
| 3rd mode | 22.25 | 27.19 | 28.78 | 30.53 | 27.18 | 28.36 | 32.24 |

On comparing the results presented in Table 1 and Table 2 it is seen that the fundamental frequency of the aluminum panel is higher compared to those for 2-layer orthotropic lamina and as a consequence the average SPL is lower for the aluminum panel. This behavior fully complies with the general understanding of mechanics where the response for a stiffer panel is lower. Whereas, for the 4-layer laminated panels the fundamental frequencies even though are higher compared to that of the aluminum panel, the average quadratic SPL for the orthotropic panels is also higher. It is to note that in all the case studies the damping ratio is considered 0.0025. This phenomenon suggests that the response for carbon fiber reinforced laminates is higher when compared with isotropic panels of the same size subjected to TBL loading.

It is further observed from the SPL results presented in Table 2 that the response in dB in the fundamental mode gets reduced when the number of layers are increased from 2 to 4 for $(0/90)_n$ and $(30/-30)_n$ lamina sequence. On the contrary for $(45/-45)_n$ lamina although the fundamental frequency for 4 – layer layup is significantly higher than 2 – layer layup suggesting a stiffer laminate, the average quadratic SPL is slightly increased for the 4 – layer layup panel.

Also, it is very interesting to note here that among the three configurations of 4 – layer laminates considered for the present study, (45/-45/45/-45) laminate shows the maximum SPL values when subjected to TBL excitation. A static deflection study is carried out for the $(\theta/-\theta/\theta/-\theta)$ angle ply laminates and (0/90/0/90) lamina by applying a distributed 1 N transverse load over the panels and the central deflection of the panel is presented in Table 3. The central deflection for the lamination sequence with $\theta = 45$ is the least suggesting it to be the stiffest laminate; whereas, the response for the same is the maximum. The extension–bending coupling provided by the antisymmetric laminate sequence along with the cross-spectrum phenomenon of the TBL model plays an important role in this not-so-normal behavior in response of antisymmetric laminates subjected to TBL excitation.

**Table 3 Central deflection ($\times\ 10^{-4}$m) of 0.5m x 0.35 m simply supported orthotropic plates subjected to a uniformly distributed load of $5.714\ \text{N/m}^2$**

| Case | Case I | Case II | Case III | Case IV | Case V | Case VI | Case VII |
| --- | --- | --- | --- | --- | --- | --- | --- |
| Deflection | --- | --- | 0.213 | --- | 0.197 | --- | 0.149 |

**5 Conclusions**

In the present research work a numerical framework based on a coupled FE-BE model is successfully implemented to obtain the structural response and resulting acoustic radiation for a baffled orthotropic lamina subjected to TBL excitation. Cholesky's decomposition technique is made used to calculate the excitation function from the PSD matrix. The major advantage of the present model over existing models is that it can be used for any structural geometry, material properties, and boundary condition. As it is known that the computational space required for BEM is less as compared to FEM, this also comes as an advantage when dealing with large acoustic domain.

It is known that the symmetric primary mode or the pumping mode contributes the highest amount of radiated SPL. The fundamental frequency of the orthotropic lamina for the lamina configuration studied in the present work although is higher as compared to the isotropic plate of the same geometrical configuration suggesting the orthotropic lamina being stiffer, the radiated average quadratic SPL is higher for the orthotropic lamina for the pumping mode subjected to TBL excitation. This observation might play a crucial role in all future development of the vehicular structure. It is also found that the present geometrical configuration $(45/-45)_n$ lamination sequence shows an irregular trend as far as radiated SPL and structural response is concerned subjected to TBL excitation.

The present numerical framework can be easily extended in the future to study TBL excited panels with TBL pressure obtained from the computational fluid dynamics model. The study can also be extended for shell panels with various configurations and also for interior acoustic problems.

# 7 Nomenclature

| | | |
|---:|:---:|:---|
| $\xi_1, \xi_3$ | = | Separation length between two points along stream-wise and cross-flow directions respectively |
| $\omega$ | = | Radial frequency |
| $\omega_n$ | = | Natural frequency |
| $\Phi_p$ | = | Single-point wall-pressure spectrum |
| $\Phi_{pp}$ | = | Cross power spectrum of wall-pressure fluctuations |
| $\Gamma$ | = | Spatial coherence function |
| $U_0$ | = | Free stream velocity |
| $U_c$ | = | Convective velocity |
| $[M]$ | = | Element mass matrix |
| $[K]$ | = | Element stiffness matrix |
| $[\rho]$ | = | Plate inertia matrix |
| $[M_s]$ | = | Structural mass matrix |
| $[C_s]$ | = | Structural damping matrix |
| $[K_s]$ | = | Structural stiffness matrix |
| $E$ | = | Modulus of elasticity |
| $G$ | = | Modulus of rigidity |
| $\nu$ | = | Poisson ratio |
| $k_s$ | = | Shear correction factor |
| $F_{tbl}$ | = | TBL induced force in time domain |
| $F(\omega)$ | = | TBL induced force in frequency domain |
| $d_s, \dot{d}_s, \ddot{d}_s$ | = | Structural displacement, velocity and acceleration respectively, in time domain |
| $d_s(\omega), \dot{d}_s(\omega), \ddot{d}_s(\omega)$ | = | Structural displacement, velocity and acceleration respectively, in frequency domain |
| $p$ | = | Sound pressure radiated by the panel excited by the TBL |
| $g$ | = | Free space Green's function |
| $k$ | = | Acoustic wave number |
| $q$ | = | Structural displacement in modal domain |
| $m$ | = | Modal mass |
| $H$ | = | Frequency response function |
| $\varphi$ | = | Mode shape |
| $P_{rad}$ | | Radiated sound power |
| $P_{ref}$ | | Reference sound power |

**Conflict of interests**

On behalf of all authors, the corresponding author states that there is no conflict of interest.

**Figure captions:**

**Fig. 1** Schematic of turbulent flow over a baffled plate

**Fig. 2** Flow chart representing TBL induced structural vibration and radiated sound power estimation using Cholesky decomposition and FE-BE coupled solver

**Fig. 3** Comparison of plate velocity PSD due to turbulent flow over baffled steel plate of dimension 0.47m × 0.37m and thickness 1.59mm with all sides clamped. Structural damping ratio is 0.0025. Flow velocity 44.7m/s.

**Fig. 4** Averaged quadratic velocity $\langle V^2 \rangle$ (in dB) of simply supported plate with $L_x$ = 0.5m, $L_y$ = 0.35 m. thickness = 2 mm subjected to a TBL flow with flow velocity of 44.7 m/s (a) Case I; Aluminum (b) Case II & III; orthotropic $(0/90)_n$ (c) Case IV & V; orthotropic $(30/-30)_n$ (d) Case VI & VII; orthotropic $(45/-45)_n$

**Fig. 5** Average radiated SPL $L_{p,rad}$ (in dB) of simply supported plate with $L_x$ = 0.5m, $L_y$ = 0.35m. thickness = 2mm subjected to a TBL flow with flow velocity of 44.7 m/s (a) Case I; Aluminum (b) Case II & III; orthotropic $(0/90)_n$ (c) Case IV & V; orthotropic $(30/-30)_n$ (d) Case VI & VII; orthotropic $(45/-45)_n$

**List of tables:**

Table 1 Frequency (in Hz) and mode numbers for simply supported rectangular panels having dimension 0.5 m x 0.35 m x 0.002 m

Table 2 Average radiated SPL (in dB) from rectangular simply supported 0.5 m x 0.35 m x 0.002 m panels placed in a baffle subjected to a flow velocity of 44.7 m/s

Table 3 Central deflection ($\times 10^{-4}$ m) of 0.5m x 0.35 m simply supported orthotropic plates subjected to a uniformly distributed load of $5.714 \, \text{N/m}^2$